\documentclass[12pt]{article}
\usepackage{epsfig}

\newcommand\pubnumber{FTUAM 01/11, IFT-UAM/CSIC 01-28 }
\newcommand\pubdate{\today}
\newcommand\hepnumber{hep-ph/0109291}
\newcommand{\nn}{\nonumber}

\newcommand{\be}{\begin{equation}}
\newcommand{\ee}{\end{equation}}
\newcommand{\bea}{\begin{eqnarray}}
\newcommand{\eea}{\end{eqnarray}}

\def\csumb{
 Departamento de F\'{\i}sica Te\'{o}rica, \\ Universidad Aut\'{o}noma
de Madrid \\ Cantoblanco, 28049 Madrid, Spain.} 
\def\support{\footnote{e-mail: maria.herrero@uam.es}}
 
\def\Title#1{\begin{center} {\Large\bf #1 } \end{center}}
\def\Author#1{\begin{center}{ \sc #1} \end{center}}
\def\Address#1{\begin{center}{ \it #1} \end{center}}

\newcommand\pubblock{\rightline{\begin{tabular}{l} \pubnumber\\
         \pubdate \\ \hepnumber \end{tabular}}}
\newenvironment{Abstract}{\begin{quotation}  }{\end{quotation}}
\newenvironment{Presented}{\begin{quotation} \begin{center} 
             Presented at the\end{center}
      \begin{center}\begin{large}}{\end{large}\end{center} \end{quotation}}
\def\Acknowledgments{\bigskip  \bigskip \begin{center}
          \large\bf Acknowledgments\end{center}}

\makeatletter
\def\section{\@startsection{section}{0}{\z@}{5.5ex plus .5ex minus
 1.5ex}{2.3ex plus .2ex}{\large\bf}}
\def\subsection{\@startsection{subsection}{1}{\z@}{3.5ex plus .5ex minus
 1.5ex}{1.3ex plus .2ex}{\normalsize\bf}}
\def\subsubsection{\@startsection{subsubsection}{2}{\z@}{-3.5ex plus
-1ex minus  -.2ex}{2.3ex plus .2ex}{\normalsize\sl}}

\renewcommand{\@makecaption}[2]{%
   \vskip 10pt
   \setbox\@tempboxa\hbox{\small #1: #2}
   \ifdim \wd\@tempboxa >\hsize     
       \small #1: #2\par          
     \else                        
       \hbox to\hsize{\hfil\box\@tempboxa\hfil}
   \fi}

 \def\citenum#1{{\def\@cite##1##2{##1}\cite{#1}}}
 
\newcount\@tempcntc
\def\@citex[#1]#2{\if@filesw\immediate\write\@auxout{\string\citation{#2}}\fi
  \@tempcnta\z@\@tempcntb\m@ne\def\@citea{}\@cite{\@for\@citeb:=#2\do
    {\@ifundefined
       {b@\@citeb}{\@citeo\@tempcntb\m@ne\@citea\def\@citea{,}{\bf ?}\@warning
       {Citation `\@citeb' on page \thepage \space undefined}}%
    {\setbox\z@\hbox{\global\@tempcntc0\csname b@\@citeb\endcsname\relax}%
     \ifnum\@tempcntc=\z@ \@citeo\@tempcntb\m@ne
       \@citea\def\@citea{,}\hbox{\csname b@\@citeb\endcsname}%
     \else
      \advance\@tempcntb\@ne
      \ifnum\@tempcntb=\@tempcntc
      \else\advance\@tempcntb\m@ne\@citeo
      \@tempcnta\@tempcntc\@tempcntb\@tempcntc\fi\fi}}\@citeo}{#1}}
\def\@citeo{\ifnum\@tempcnta>\@tempcntb\else\@citea\def\@citea{,}%
  \ifnum\@tempcnta=\@tempcntb\the\@tempcnta\else
  {\advance\@tempcnta\@ne\ifnum\@tempcnta=\@tempcntb \else\def\@citea{--}\fi
    \advance\@tempcnta\m@ne\the\@tempcnta\@citea\the\@tempcntb}\fi\fi}
\makeatother

\def\beq{\begin{equation}}
\def\eeq#1{\label{#1}\end{equation}}
\def\eeqn{\end{equation}}

\newenvironment{Eqnarray}%
   {\arraycolsep 0.14em\begin{eqnarray}}{\end{eqnarray}}
\def\beqa{\begin{Eqnarray}}
\def\eeqa#1{\label{#1}\end{Eqnarray}}
\def\eeqan{\end{Eqnarray}}

\let\bar=\overbar

\def\Dslash{\not{\hbox{\kern-4pt $D$}}}
\def\dslash{\not{\hbox{\kern-2pt $\del$}}}

\def\ee{e^+e^-}

\def\msb{{\bar{\ssstyle M \kern -1pt S}}}

\def\lsim{\mathrel{\raise.3ex\hbox{$<$\kern-.75em\lower1ex\hbox{$\sim$}}}}
\def\gsim{\mathrel{\raise.3ex\hbox{$>$\kern-.75em\lower1ex\hbox{$\sim$}}}}

\begin{document}
\begin{titlepage}
\begin{flushright}
\pubblock
\end{flushright}

\vfill
\def\thefootnote{\fnsymbol{footnote}}
\Title{Indirect heavy SUSY signals \\[5pt] in Higgs and 
top decays}
\vfill
\Author{Mar\'{\i}a J. Herrero\support}
\Address{\csumb}
\vfill
\begin{Abstract}
We summarize the recent results on the supersymmetric QCD radiative 
corrections, at the one-loop
level, in Higgs and top quark decays, in the 
context of the Minimal Supersymmetric Standard Model and in the 
decoupling limit of very heavy SUSY particles. Special attention is devoted
to the particular decays $h^0\rightarrow b \bar b$, $H^+\rightarrow t \bar b$,
$H^0\rightarrow b \bar b$, $A^0\rightarrow b \bar b$ and 
$t \rightarrow H^+\bar b$ where the radiative corrections from heavy squarks 
and heavy gluinos do not decouple and are enhanced at large $\tan \beta$. Some 
interesting phenomenological consequences are also briefly summarized.
\end{Abstract}
\vfill
\begin{Presented}
XXIX International Meeting on Fundamental Physics\\ 
Seminar in honor of Prof. F.J. Yndur\'ain \\[4pt]
Sitges, Barcelona, Spain, 5-9 February, 2001
\end{Presented}
\vfill
\end{titlepage}
\def\thefootnote{\arabic{footnote}}
\setcounter{footnote}{0}
%
\section{Introduction}
 One of the most challenging goals of the next generation colliders is 
 the discovery of supersymmetric (SUSY) particles and the study of their rich 
 associated phenomenology. In the context of the Minimal Supersymmetric Standard
 Model (MSSM)~\cite{MSSM}, the SUSY physical spectrum consists of squarks, 
 $\tilde q_1,\tilde q_2$ ($q=u,d,s,c,b,t$), sleptons, 
 $\tilde l_1,\tilde l_2$ ($l=e,\mu,\tau$), sneutrinos, 
 $\tilde \nu_l$ ($l=e,\mu,\tau$), gluinos,  ${\tilde g}_a$ ($a=1,..,8$),
 charginos, $\tilde \chi^{\pm}_{i}$ ($i=1,2$) and neutralinos, 
 $\tilde \chi^{o}_{j}$ ($j=1,..,4$). In addition, the 
 MSSM incorporates an extended Higgs sector with two Higgs doublets that lead 
 to five physical Higgs boson particles: two CP-even scalars, $h^0$ and $H^0$,
 one CP-odd scalar, $A^0$, and two charged scalars, 
 $H^{\pm}$~\cite{2HDM}. None of these
 non-standard particles have been discovered yet, and the present experiments 
 have placed some lower mass limits. From the last published Review of Particle
 Physics~\cite{PDG2000} the following $95\% CL$ limits have been extracted: 
 $M_{\tilde q}>250\,GeV$ ($q\neq t,b$), $M_{\tilde t}>86.4\,GeV$, 
 $M_{\tilde b}>75\,GeV$, $M_{\tilde e}>87.1\,GeV$,
$M_{\tilde \mu}>82.3\,GeV$, $M_{\tilde \tau}>81.0\,GeV$, 
$M_{\tilde \nu}>43.1\,GeV$, $M_{\tilde g}>190\,GeV$, 
$M_{\tilde \chi^{\pm}_{1}}>67.7\,GeV$, $M_{\tilde \chi^{o}_{1}}>32.5\,GeV$,
$M_{h^0}>82.6\,GeV$, $M_{A^0}>84.1\, GeV$, $M_{H^\pm}>69.0 \,GeV$.
Most of these limits are dependent on particular assumptions on the MSSM
parameters and have been improved after that Review of Particle 
Physics~\footnote{The 2001 summer conferences 
(see for instance ref.\cite{updated}) have 
announced updated $95\% CL$ limits:
 $M_{\tilde q}>300\,GeV$ ($q\neq t,b$), $M_{\tilde t}>95\,GeV$, 
 $M_{\tilde b}>92\,GeV$, $M_{\tilde e}>99\,GeV$,
$M_{\tilde \mu}>95\,GeV$, $M_{\tilde \tau}>80.0\,GeV$, 
$M_{\tilde \nu}>43.1\,GeV$, $M_{\tilde g}>195\,GeV$, 
$M_{\tilde \chi^{\pm}_{1}}>101\,GeV$, $M_{\tilde \chi^{o}_{1}}>45.6\,GeV$,
$M_{h^0}>91.0\,GeV$, $M_{A^0}>91.9\, GeV$, $M_{H^\pm}>78.6 \,GeV$.
   }  
        
The absence of SUSY particles in the present colliders still leaves the possibility
 that these particles manifest at
energies larger than the present explored energies, being these latter typically
of the
order of the electroweak scale, $M_{EW}\sim {\cal O}(245\,GeV)$. 
In this work we will assume that the 
whole spectrum of genuine SUSY particles is heavy,
 such that they can not be produced directly at the present or next
generation colliders, 
and different strategies based on indirect searches must be performed.

The study of radiative corrections from SUSY particles 
to Standard Model (SM) couplings and SM observables may provide crucial 
clues in this search of indirect SUSY signals if the SUSY masses 
are beyond the 
reach of present and planned accelerators~\cite{proceedings}. 
In particular, if a light Higgs boson, $h^0$,
were discovered 
in the mass range predicted by the MSSM, 
$M_{h^0}\le 135\, GeV$~\cite{loopmassMSSM}, 
but the SUSY particles were not found, a precise 
measurement of Higgs couplings to SM particles, which are sensitive to 
radiative corrections, could provide
indirect information about the existence of SUSY in Nature and even serve to
infer the SUSY particle masses. For example, one could
conclude (in the context of the MSSM)
whether the data favor a SUSY spectrum below the 1 TeV energy scale.
Similar studies can be performed by considering some relevant observables as,
for instance,  
the partial widths of the top quark and Higgs boson $h^0$ decays into 
SM particles, and by comparing 
their predictions in the MSSM and the SM. Furthermore, in case the 
extra Higgs
boson particles, $H^0$, $A^0$ and $H^{\pm}$, be accessible as well to the 
present or
future colliders, one can also study the sensitivity to a heavy SUSY spectrum 
via the SUSY radiative corrections to their relevant partial decay widths and 
compare their predictions with those of a more general two-Higgs-doublets model
(2HDM). There is an extensive literature on SUSY radiative corrections to 
the decays of Higgs particles and the top quark, in the context of the MSSM. 
We refer the reader to 
refs.~\cite{Dabelstein,cjs,solaHTB,EberlHTB,solaEW,topQCD,topEW,cmwpheno,polonsky,kolda,CarenaDavid,EberlTodos,HaberTemes,ourHtb,RADCOR2000,siannah,HaberLC,dobado,curiel}
for a selection of works closely related with the subject of this contribution.    

In this review we will consider the dominant SUSY radiative corrections to the 
relevant Higgs boson particles and top quark decays, at the one-loop level,
that come from the 
SUSY-QCD (SQCD) sector, that is from squarks and gluinos, and study their behavior  
in the limit 
where the SUSY particles are very heavy  as compared to the 
electroweak scale, $M_{SUSY}\gg M_{EW}$, where $M_{SUSY}$ represents generically 
the masses of the SUSY particles. This situation corresponds to the decoupling 
of SUSY particles from the rest of the MSSM spectrum, namely, the SM particles 
and the MSSM Higgs sector. Special
attention will be devoted to the particular decays $h^0\rightarrow b \bar b$, $H^+\rightarrow t \bar b$,
$H^0\rightarrow b \bar b$, $A^0\rightarrow b \bar b$ and 
$t \rightarrow H^+\bar b$ where the SUSY-QCD radiative corrections from heavy squarks 
and heavy gluinos turn out to be non-decoupling from the low energy physics 
and, furthermore, they are enhanced at large $\tan \beta$. 
This non-decoupling behavior is genuine of Higgs 
sector physics and has not been found yet in other MSSM sectors, as for instance in
electroweak gauge bosons physics~\cite{TesisS}. It has not been seen in
the $h^0$ self-couplings either~\cite{HollikSiannah}. The interest of these
non-decoupling effects is that they can produce sizeable contributions 
to the Higgs particles and top quark  partial decay widths 
for large enough $\tan \beta$ values and, 
therefore, can provide indirect SUSY signals, even for very 
heavy squarks and gluinos. We will briefly comment on some recently proposed 
observables that are defined as ratios of Higgs branching ratios into quarks 
divided by
the corresponding Higgs branching ratios into leptons, and that turn out to be  
the most sensitive to these SUSY non-decoupling contributions. We will devote 
as well some comments to the alternative plausible possibility where, not just
the SUSY particles but also the extra Higgses are heavy.  This limiting
situation
occurs when both $M_{SUSY}$ and the CP-odd Higgs boson mass $M_A$ are 
larger than $M_{EW}$,  and the decoupling of all non-standard particles 
from the SM physics is expected. In this limit, the lightest Higgs boson, 
$h^0$, resembles the SM Higgs particle, and a distinction between the MSSM and
the SM will be very difficult~\cite{decoupling}.

We present here just a summary of the main results and refer the reader 
to refs.~\cite{HaberTemes,ourHtb,RADCOR2000,dobado,curiel,TesisS} for more details.

\section{Decoupling limit in the SUSY-QCD sector}  

In this section we shortly review the SUSY-QCD sector of the MSSM, 
consisting of squarks and 
gluinos, and define 
the decoupling limit, where these SUSY particles are much heavier than the electroweak
scale.  We will concentrate on the third-generation squarks since they
provide the dominant radiative corrections to the Higgs particles and top quark
decays.

The sbottom and stop mass matrices are given in terms of the MSSM parameters 
respectively by:


$${\hat M}_{\tilde{b}}^2 =\left(\begin{array}{cc}  
 { {M_{{\scriptscriptstyle {\tilde Q}}}^{2}}+ m_{b}^{2} - 
 M_{Z}^{2} (\frac{1}{2}+Q_b   
s_{{\scriptscriptstyle W}}^{2}) \cos{2\beta}} & 
{m_{\scriptscriptstyle b}( {A_b}- {\mu} \tan{\beta})}  
\\ {m_{\scriptscriptstyle b}( {A_b}- {\mu} \tan{\beta})} &
{  {M_{{\scriptscriptstyle {\tilde D}}}^{2}} +    
m_{b}^{2} + M_{Z}^{2} Q_b s_{{\scriptscriptstyle W}}^{2} \cos{2\beta}}   
\end{array} \right), $$

and
   
$${\hat M}_{\tilde{t}}^2 =\left(\begin{array}{cc}  
 { {M_{{\scriptscriptstyle {\tilde Q}}}^{2}}+ m_{t}^{2} +
  M_{Z}^{2} (\frac{1}{2}-Q_t   
s_{{\scriptscriptstyle W}}^{2}) \cos{2\beta}} & 
{m_{\scriptscriptstyle t}( {A_t}- {\mu} \cot{\beta})}  
\\ {m_{\scriptscriptstyle t}( {A_t}- {\mu} \cot{\beta})} &
{  {M_{{\scriptscriptstyle {\tilde U}}}^{2}} +   
m_{t}^{2} + M_{Z}^{2} Q_t s_{{\scriptscriptstyle W}}^{2} \cos{2\beta}}  
\end{array} \right), $$
where $M_{{\scriptscriptstyle {\tilde Q}}}$, 
$M_{{\scriptscriptstyle {\tilde U}}}$, 
$M_{{\scriptscriptstyle {\tilde D}}}$, are the soft-SUSY-breaking mass
parameters for the squark doublet ${\tilde q_L}$ and the squark singlets
${\tilde t_R}$ and ${\tilde b_R}$, respectively.  $A_{t,b}$ are the stop and
sbottom  soft-SUSY-breaking trilinear couplings, respectively. 
The 
$\mu$-parameter is the SUSY-preserving bilinear Higgs coupling. The ratio 
of the two Higgs vacuum spectation values is given by
$\tan{\beta}=v_2/v_1$. $M_Z$, $m_t$ and $m_b$ are the standard $Z$ boson, top 
quark and bottom quark masses. $Q_t=2/3$, $Q_b=-1/3$ and 
$s_{{\scriptscriptstyle W}}=\sin \theta_{{\scriptscriptstyle W}}$.  

The physical squared masses of the sbottoms, $M_{\tilde b_{1,2}}^2$, and stops, 
$M_{\tilde t_{1,2}}^2$, are the eigenvalues of the previous mass matrices,  
${\hat M}_{\tilde{b}}^2$ and ${\hat M}_{\tilde{t}}^2$, respectively.
The mass of the gluinos, ${\tilde g}_a$ ($a=1,..,8$), is given by the 
soft-SUSY-breaking Majorana mass $M_{\tilde g}$.    

In order to get heavy squarks and heavy gluinos, we need to choose 
properly the soft-SUSY-breaking parameters and the $\mu$-parameter. Since here 
we are interested in the limiting situation where the whole 
SUSY spectrum is heavier than the electroweak scale, we have made
the simplest assumption for the soft breaking squark mass parameters, trilinear
terms, $\mu$-parameter and gluino mass (see refs.~\cite{HaberTemes,ourHtb} for more 
details),

$$M_{SUSY} \sim M_{{\scriptscriptstyle {\tilde Q}}} 
\sim M_{{\scriptscriptstyle {\tilde D}}}
\sim M_{{\scriptscriptstyle {\tilde U}}}
\sim M_{\tilde g} 
\sim \mu  \sim A_{b} \sim A_{t} \gg M_{EW}, $$ 
where $M_{SUSY}$ represents generically a common SUSY large mass scale and 
the symbol '$\sim$' means quantities of the same order of magnitude but 
not necessarily equal. In the following of this work, this general condition 
will be referred to as 'large SUSY mass limit'. 

We have considered two extreme cases, maximal and minimal squark mixing,  
 which, for the previous 'large SUSY mass limit', imply certain constraints 
 on the squark mass differences.
 Thus, given the generic mass matrix,   

\vspace{-0.3cm}

$${\hat M}^2_{\tilde q} \equiv\left(\begin{array}{cc}  
 { M_{q,L}^2} & 
{m_q X_q}  
\\ {m_q X_q} &
{M_{q,R}^2}   
\end{array} \right), $$
the two limiting cases are reached by choosing the relative size of 
$M_{q,L,R}$ and $X_q$ as follows,

A.-Close to maximal mixing:
$\theta_{\tilde q} \sim \pm 45^{\circ}$
 
\vspace{-0.2cm}

$$|M_{q,L}^2-M_{q,R}^2|\ll m_q X_q \Rightarrow 
|M_{\tilde q_1}^2-M_{\tilde q_2}^2|\ll 
|M_{\tilde q_1}^2+M_{\tilde q_2}^2|$$

B.-Close to minimal mixing:
 $\theta_{\tilde q} \sim 0^{\circ}$
 
\vspace{-0.2cm}

$$|M_{q,L}^2-M_{q,R}^2|\gg m_q X_q \Rightarrow 
|M_{\tilde q_1}^2-M_{\tilde q_2}^2|\sim
{\cal O}|M_{\tilde q_1}^2+M_{\tilde q_2}^2|$$
where we have included the corresponding implications for the physical 
squark mass differences.


\section{Decoupling  limit in the Higgs sector}

 The decoupling limit in the Higgs sector of the MSSM was
 first studied in ref.~\cite{decoupling}. In short, it is defined by considering
 the CP-odd Higgs mass much larger than the electroweak mass, $M_A \gg M_Z$,
 and leads to a particular spectrum in the Higgs sector 
 with very heavy $H^0$, $H^{\pm}$ and $A^0$ bosons, and a light $h^0$ boson.
 For a review of the MSSM Higgs sector, see ref.~\cite{2HDM}.
  
 At tree level, if $M_A \gg M_Z$, the Higgs masses are, 

\vspace{0.1cm}
 \hspace{2.5cm}
 $ M_{h^0} \simeq M_{H^{\pm}} \simeq M_A \gg M_Z 
\,\,\,,\,\, M_{h^0} \simeq M_Z |cos2\beta|\,.$\\
That is, at tree-level there exists  a  CP-even Higgs, $h^0$, 
lighter than the $Z$ boson.
 
\begin{table}[hbt]
\renewcommand{\arraystretch}{1.5}
\begin{center}
\begin{tabular}{|lc||ccc|} \hline
\multicolumn{2}{|c||}{$\phi$} & $g_{\phi \bar tt}$ & $g_{\phi \bar bb}$ & 
 $g_{\phi VV}$ \\
\hline \hline
SM~ & $H$ & 1 & 1 & 1 \\ \hline
MSSM~ & $h^o$ & $ {\cos\alpha/\sin\beta}$ & $ {-\sin\alpha/\cos\beta}$ &
$ {\sin(\beta-\alpha)}$ \\
& $H^o$ & $\sin\alpha/\sin\beta$ & $\cos\alpha/\cos\beta$ &
$\cos(\beta-\alpha)$ \\
& $A^o$ & $ 1/\tan\beta$ & $\tan\beta$ & 0 \\ \hline
\end{tabular}
\renewcommand{\arraystretch}{1.2}
\caption{\label{hcoup} Higgs couplings in the MSSM normalized to SM 
couplings, in terms of $\beta$ and the mixing angle of the neutral 
CP-even Higgs sector $\alpha$. Here $V=Z,W$.} 
\end{center}
\end{table}

 Concerning the neutral Higgs couplings, their tree-level values in the 
 MSSM, normalized 
 to SM couplings and for arbitrary $M_A$, are given in table~\ref{hcoup}. 
  

Notice that by expanding in inverse powers of $M_A$, we get:

\vspace{0.1cm}

\hspace{2cm} $ {\frac{\cos\alpha}{\sin\beta}}\simeq  1 + 
\cal{O}$$(M_Z^2/M_A^2)$$\,,\,$
$ {-\frac{\sin\alpha}{\cos\beta}}\simeq 1 + 
\cal{O}$$(M_Z^2/M_A^2)$

\vspace{0.3cm}

\hspace{3.5cm} $ {\sin(\beta - \alpha)} \simeq 1 + 
\cal{O}$$(M_Z^4/M_A^4)$.
\vspace{0.1cm}

Therefore, the $h^0$ tree-level couplings in the decoupling limit, 
$M_A \gg M_Z$,
tend to their SM values, as expected.

Beyond tree level, it has been shown~\cite{decoupling} that, in this same decoupling 
limit, the Higgs masses keep a similar pattern as at tree level, that is, 
very heavy $H^0$, $H^{\pm}$ and $A^0$ bosons, and a light $h^0$ boson. The
particular values of their masses depend of course on the MSSM parameters, 
but for 
$M_A \gg M_Z$,  

\vspace{0.1cm}

\hspace{2cm}{  $M_{h^0} \simeq M_{H^{\pm}} \simeq M_A \gg M_Z\,,$} and  
$M_{h^0} \le  135 \, GeV\,.$

In this work we will go beyond tree level and study    
  the decoupling behavior of heavy  
 SUSY particles and heavy Higgses, at one-loop level.  
  
\section{Decoupling in low-energy electroweak gauge boson physics}

 It has been shown that all one-loop corrections to low-energy 
  electroweak gauge boson physics involving SUSY particles and extra Higgs
  bosons 
  decouple in the limit of large sparticle masses and large $M_A$~\cite{TesisS}.
 The formal proof of this decoupling involves the  
  computation of the effective action for electroweak gauge bosons 
  $\Gamma_{eff}[A,Z,W^\pm]$ by explicit integration, in the path integral,
  of all the heavy sparticles and heavy extra Higgs bosons, 
$\tilde f(\tilde f=\tilde q,\tilde l,\tilde {\nu_l}),
 \tilde{\chi}^{{\scriptscriptstyle \pm}}_{{\scriptscriptstyle
i}},\tilde{\chi}^{{\scriptscriptstyle o}}_{{\scriptscriptstyle
j}},H (H= H^{\pm},H^0,A^0)$ at the one-loop level,
$${\rm e}^{i\Gamma_{eff}[V]}=\int 
 [{\rm d}\tilde f]\,
 [{\rm d}\tilde f^*]\,
 [{\rm d}\tilde{\chi}^{{\scriptscriptstyle +}}]\,
 [{\rm d}\bar{\tilde{\chi}}^{{\scriptscriptstyle +}}]\,
 [{\rm d}\tilde{\chi}^{{\scriptscriptstyle 0}}]\,
 [{\rm d} H ]\,
 {\rm e}^{i \Gamma_{\rm MSSM}[V,\tilde f,
 \tilde{\chi}^{{\scriptscriptstyle +}},
 \tilde{\chi}^{{\scriptscriptstyle 0}},H]}\,;\, V=A,Z,W^\pm .$$
\begin{figure}
\hspace{-1.5cm}\epsfig{file=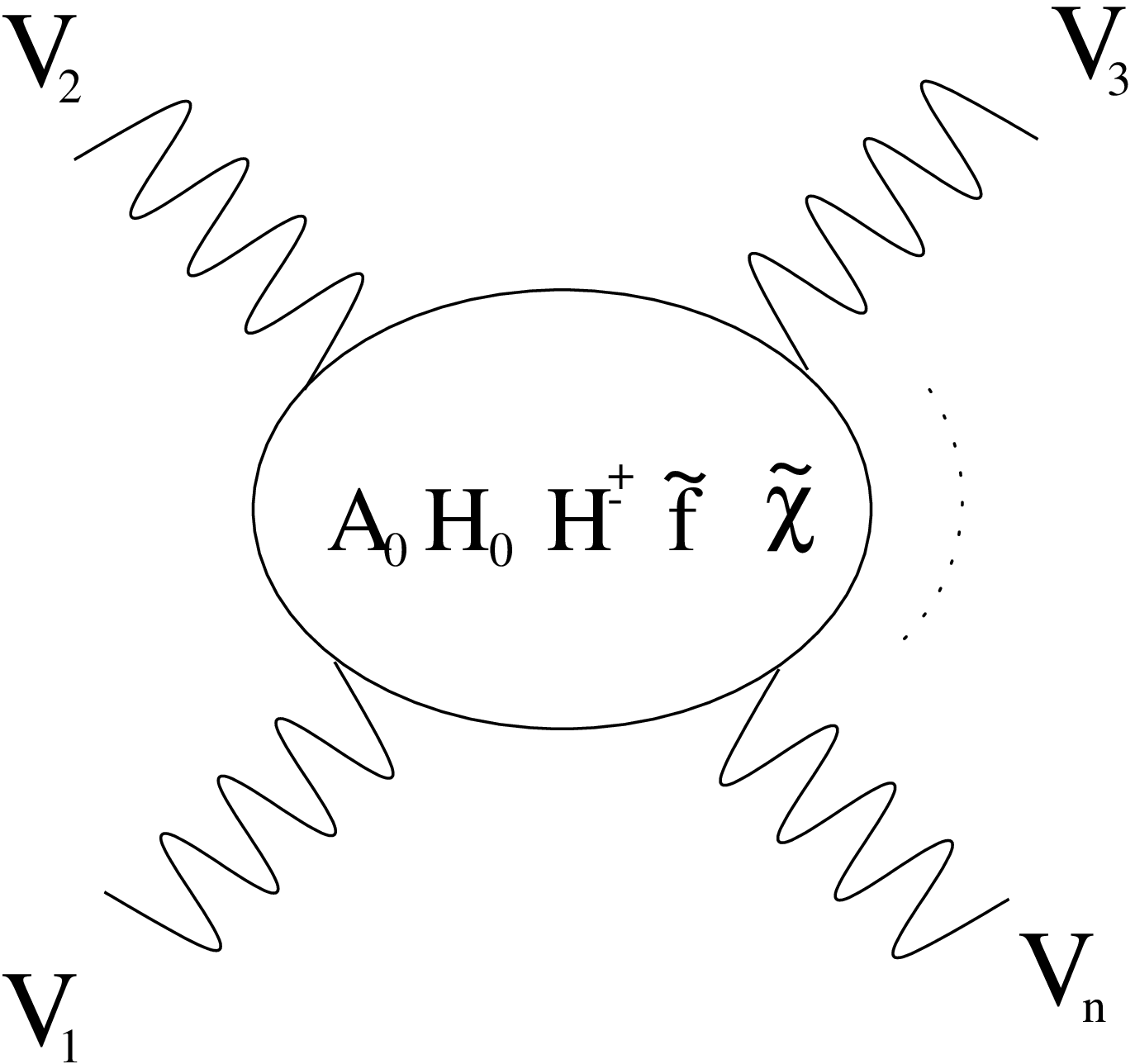,width=4.5cm,height=4.5cm}
\hspace{1cm}\epsfig{file=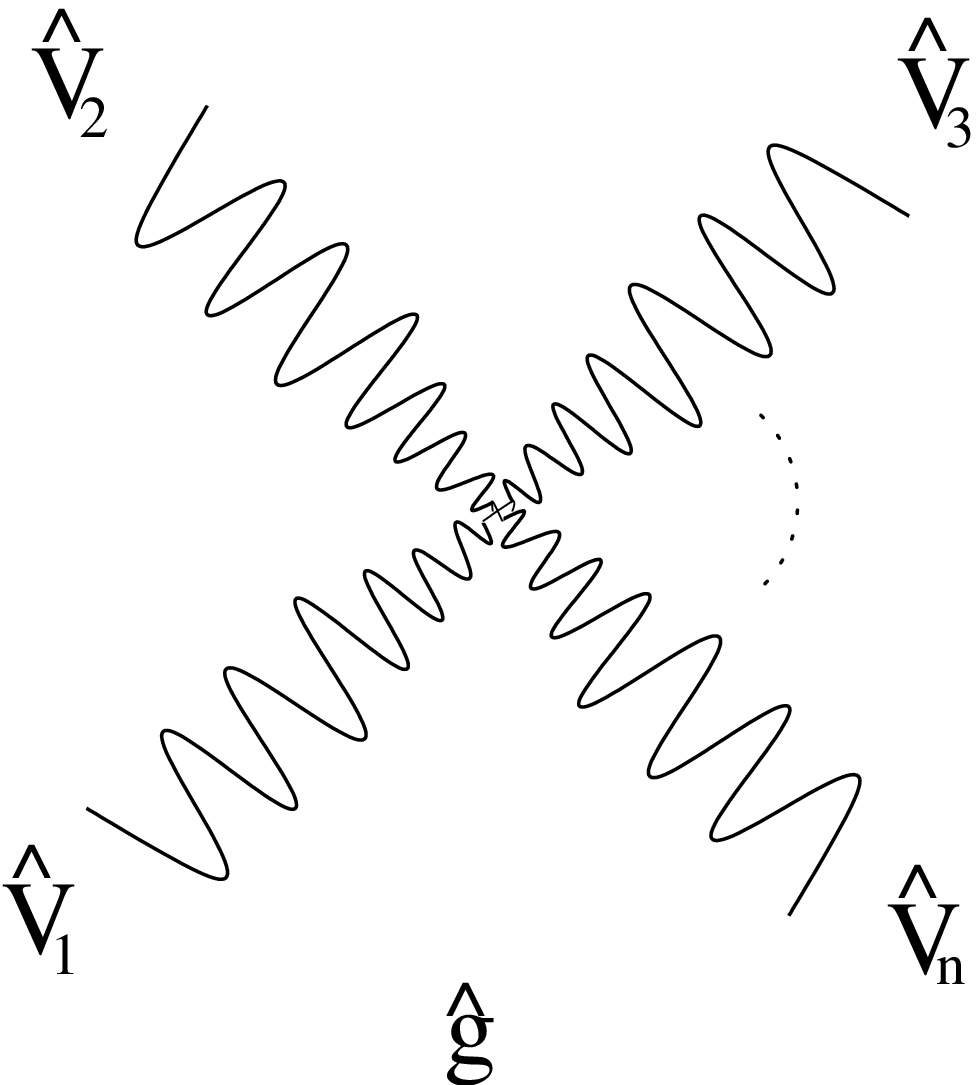,width=4cm,height=4cm}

\vspace{-3cm}

\hspace{3cm}\Huge {$\rightarrow$}\\

\vspace{-2cm}
\hspace*{8cm}\LARGE {+} \large {$\, \, \mathcal{O}$$\left(\left(\frac{p}{M_{SUSY}}\right)^m,\left(\frac{M_{EW}}{M_{SUSY}}\right)^r,\left(\frac{\Delta M_{SUSY}}{M_{SUSY}}\right)^t \right)$}
\vspace{2cm}
\caption[0]{Decoupling of heavy sparticles and heavy Higgses in 
the n-point functions of the electroweak gauge bosons $V=A,Z,W^\pm$.}
\label{fig:fig0}
\end{figure} 
 The proof involves, in addition, a large sparticle masses and large $M_A$ 
 expansion of 
 $\Gamma_{eff}[A,Z,W^\pm]$ that is valid in the heavy SUSY masses
 limit,
 $M_{SUSY}\sim
m_{\scriptstyle \tilde{f}},\,
m_{{\scriptstyle {\tilde{\chi}^{{\scriptscriptstyle \pm}}_{{\scriptscriptstyle
i}}}}},\,m_{{\scriptstyle {\tilde{\chi}^{{\scriptscriptstyle o}}_{{\scriptscriptstyle
j}}}}},\,M_A \gg M_{EW}$, with 
$|\tilde{m}_{\scriptscriptstyle i}^{2}-
\tilde{m}_{\scriptscriptstyle j}^{2}| \ll
|\tilde{m}_{\scriptscriptstyle i}^{2}+
\tilde{m}_{\scriptscriptstyle j}^{2}|\,{\scriptstyle \forall i,j}$.
 The result of this expansion is shown schematically in Fig.~\ref{fig:fig0}.  

All the effects from heavy SUSY particles and extra Higgses in the
effective action for electroweak
gauge bosons, or equivalently in the n-point functions, 
are either absorbed into redefinitions of the electroweak 
parameters (represented generically by ${\hat g}$ in Fig.~\ref{fig:fig0}) 
and external wave functions, (${\hat V_1}$,..${\hat V_n}$), or else they are 
supressed by inverse powers of 
the large masses. Therefore, in the asymptotic limit of an infinitely heavy SUSY
spectrum, these effects decouple in the physical observables 
involving external electroweak gauge bosons. This decoupling is referred to as   
decoupling {\it a la} Appelquist Carazzone~\cite{ac}.

\section{SUSY-QCD corrections to $h^o \rightarrow \bar bb$ in the decoupling
limit}

In this section we study the SUSY-QCD corrections to the partial 
decay width $\Gamma(h^o \rightarrow \bar bb)$ at the one-loop level and to leading
order in perturbative QCD, that is ${\cal O}$($\alpha_S$). We will then explore
the decoupling behaviour of these corrections for large SUSY masses, $M_{SUSY}$,
and/or large $M_A$. Both numerical and analytical results will be 
presented~\cite{HaberTemes}.

For the $h^0$ mass range predicted by the MSSM, the decay channel
$h^o \rightarrow \bar bb$ is by far the dominant one and the precise 
value of its branching ratio 
will be crucial for the $h^0$ final experimental reach at the Tevatron.
  
Among the various contributions to this decay width, the QCD 
 corrections are known to be the dominant ones. At the one-loop level and to order 
$\alpha_S$ these can be written as,  
$$\Gamma_1(h^o \to b \bar b) \equiv \Gamma_0(h^o \to b \bar b)
        (1 + 2   {\Delta_{QCD}} + 2    {\Delta_{SQCD}}),$$
where, $\Gamma_0(h^o \to b \bar b)$ is the tree level width, $\Delta_{QCD}$ 
is the one-loop contribution from standard QCD  and $\Delta_{SQCD}$ is the one-loop
contribution from the SUSY-QCD sector of the MSSM. The factor $2$ is just a convention. The QCD correction, 
$\Delta_{QCD}$, gives a $\sim  50 \%$  reduction in the
 $\Gamma (h^o \rightarrow b \bar b)$ decay rate for $M_{h^0}$ in its MSSM 
 range~\cite{qcd}. 
  This correction has the same form in the MSSM as in the SM, so that
 it gives no information in distinguishing the MSSM from the SM. 
 The SQCD correction, $\Delta_{SQCD}$, was first computed in the
  on-shell scheme by using a
  diagrammatic approach in ref.~\cite{Dabelstein} and later studied 
in more detail in~\cite{cjs}. The SQCD corrections to the $h^0b\bar b$
  coupling  were
  also computed in an effective Lagrangian approach in 
  ref.~\cite{cmwpheno}, using the SUSY contributions to
  the $b$-quark self energy~\cite{hrs-h-copw,pbmz}
  and neglecting terms suppressed by inverse powers of
  SUSY masses. Radiatively induced Higgs-fermion-fermion couplings in supersymmetric theories were also studied in ref~\cite{polonsky}. The size of the 
 SQCD 
 correction, $\Delta_{SQCD}$, 
     and the QCD correction, $\Delta_{QCD}$, are comparable 
  for a wide window of the MSSM parameter space. In some regions of the MSSM parameter space, the SQCD corrections
  become so large that it is important to take into account higher-order
  corrections.  The two-loop SQCD corrections have been studied in a
  diagrammatic approach in
  ref.~\cite{hhw}. A higher-order analysis 
   has also been carried
  out in refs.~\cite{CarenaDavid,EberlTodos} by
performing a resummation of the leading $\tan\beta$ contributions to all orders of perturbation
theory and by using an effective Lagrangian approach. However, this
resummation is not important in our present work because
we are interested in the decoupling limit, in which the one-loop
corrections to the $h^0 b\bar b$ coupling are small enough. Thus, for the present
analysis we will just keep the one-loop corrections.
\begin{figure}[h]
\begin{center}
\epsfig{file=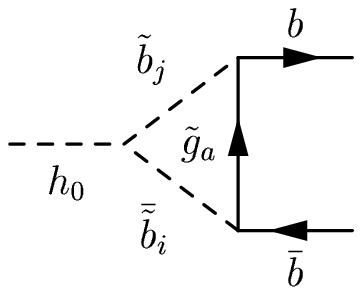}~~
\epsfig{file=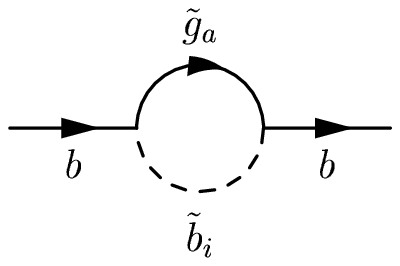}
\caption[0]{One-loop SUSY diagrams contributing to ${\cal O}$($\alpha_S$) to 
$h^o \to b \bar b$ decay}
\label{fig:fig1}
\end{center}
\end{figure} 

To one-loop and ${\cal O}$($\alpha_S$) there are two type of diagrams, shown 
in Fig.~\ref{fig:fig1}, that
contribute to $$\Delta_{SQCD}=\Delta_{SQCD}^{\rm loops}+
\Delta_{SQCD}^{\rm CT}\,.$$ 

The triangle diagram, with exchange of sbottoms and gluinos, 
contributes to $\Delta_{SQCD}^{\rm loops}$, whereas the bottom 
self-energy diagram contributes to the counter-terms part 
$\Delta_{SQCD}^{\rm CT}$. The exact results in the on-shell scheme are
summarized by,

\begin{eqnarray}
          & {\Delta _{SQCD}^{\rm loops}}=  
          \frac{\alpha_s}{3 \pi} \Bigl\{ \Bigl[ \frac{2 M_Z^2}{m_b} 
        \frac{\cos\beta \sin(\alpha + \beta)}{\sin\alpha}
        (I_3^b \cos^2\theta_{\tilde b} - Q_b s^2_W \cos 2 \theta_{\tilde b})
        + 2 m_b +   {Y_b} \sin 2 \theta_{\tilde b} \Bigr]  \nonumber \\
        & 
        \times \Bigl[ m_b   {C_{11}} +
          {M_{\tilde g}} \sin 2 \theta_{\tilde b}   {C_0} \Bigr]
        (m_b^2,M_{h^0}^2,m_b^2;
         {M_{\tilde g}^2}, {M_{\tilde b_1}^2},
                {M_{\tilde b_1}^2})
                \nonumber \\
        & 
        + \Bigl[ \frac{2 M_Z^2}{m_b} 
        \frac{\cos\beta \sin(\alpha + \beta)}{\sin\alpha}
        (I_3^b \sin^2 \theta_{\tilde b} + 
        Q_b s^2_W \cos 2 \theta_{\tilde b})
        + 2 m_b -   {Y_b} \sin 2 \theta_{\tilde b} \Bigr] 
        \nonumber \\
        & 
        \times \Bigl[ m_b   {C_{11}} - 
         {M_{\tilde g}} \sin 2 \theta_{\tilde b}   {C_0} \Bigr]
        (m_b^2,M_{h^0}^2,m_b^2;
         {M_{\tilde g}^2}, {M_{\tilde b_2}^2},
         {M_{\tilde b_2}^2})
                \nonumber \\
        & 
        + \Bigl[ -\frac{M_Z^2}{m_b} 
        \frac{\cos\beta \sin(\alpha + \beta)}{\sin\alpha}
        (I_3^b - 2 Q_b s^2_W) \sin 2 \theta_{\tilde b}
        + Y_b \cos 2 \theta_{\tilde b} \Bigr]  \nonumber \\
        & 
        \times \Bigl[ 2 { M_{\tilde g}} \cos 2 
        \theta_{\tilde b}   {C_0} 
        (m_b^2,M_{h^0}^2,m_b^2; {M_{\tilde g}^2},
         {M_{\tilde b_1}^2}, {M_{\tilde b_2}^2})
        \Bigr] \Bigr\}\,, \nonumber
        \label{eqn:fulldgvertex}
\end{eqnarray}
\begin{eqnarray}
        & {\Delta_{SQCD}^{\rm CT}}
        = -  \frac{\alpha_s}{3\pi} \Bigl\{ \frac{ {M_{\tilde g}}}{m_b} 
        \sin 2 \theta_{\tilde b}
        \Bigl[  {B_0} (m_b^2; {M_{\tilde g}^2},
         {M_{\tilde b_1}^2})
        -   {B_0} (m_b^2; {M_{\tilde g}^2},
          {M_{\tilde b_2}^2}) \Bigr]  \nonumber \\
        & 
        - 2 m_b^2 \Bigl[  {B_1^{\prime}}(m_b^2;
         {M_{\tilde g}^2}, {M_{\tilde b_1}^2})
        +   {B_1^{\prime}}(m_b^2;
          {M_{\tilde g}^2}, {M_{\tilde b_2}^2}) \Bigr]
        \nonumber \\
        & 
        - 2 m_b  {M_{\tilde g}} \sin 2 \theta_{\tilde b}
        \Bigl[  {B_0^{\prime}}(m_b^2;
         {M_{\tilde g}^2}, {M_{\tilde b_1}^2})
        -  {B_0^{\prime}}(m_b^2; {M_{\tilde g}^2},
         {M_{\tilde b_2}^2}) \Bigr]
        \Bigr\} \,,\nonumber
        \label{eqn:fulldgwfct}
\end{eqnarray}
where we have used the standard notation for masses, couplings and mixing
angles, and  we have followed the definitions and conventions for the 
one-loop integrals $B_0$, $B_0'$, $B_1'$, $C_0$ and $C_{11}$ of 
ref.~\cite{Hollik}. Notice that 

\vspace{-0.4cm}

$$ Y_b \equiv  {A_b} + { \mu} \cot\alpha $$  
appears in the 
 $h^o \tilde b_R \tilde b_L$ coupling and, therefore, it  
  will be responsible for sizeable contributions in the large $A_b$ and/or 
  $\mu$ limit. Our results agree with those of refs.~\cite{Dabelstein,cjs}.
   
 In order to compute  $\Delta_{SQCD}$ in the decoupling limit of very heavy 
 sbottoms and gluinos, we have considered the previously announced assumption 
 for the MSSM parameters, i.e, the 'large SUSY mass limit'. 
  We have performed a systematic expansion of the one-loop integrals and the
 mixing angle $\theta_{\tilde b}$ in inverse powers of the large SUSY mass
 parameters. The resulting formulas of these expansions can be found in 
 ref.~\cite{HaberTemes}. Thus, by defining 
 $$\tilde M_S^2 \equiv 
\frac{1}{2}(M_{\tilde b_1}^2 + M_{\tilde b_2}^2)\,\,,
   \,\,R \equiv \frac{M_{\tilde g}}{\tilde M_S}\,\,,\,\,
   X_b \equiv A_b-\mu \tan \beta  $$
   and including terms up to 
${\cal O}(M_{Z,h^o}^2/{\tilde M_S^2})$ in the expansion, 
we get the following result
for the maximal mixing case, $\theta_{\tilde b}\sim \pm 45 ^{\circ}$:
\begin{eqnarray}
        &  {\Delta_{SQCD}}
        = \frac{\alpha_s}{3\pi} 
        \left\{ \frac{-  {\mu M_{\tilde g}}}{  {\tilde M_S^2}}
        \left( \tan\beta + \cot\alpha \right)
        f_1(R) 
        - \frac{  {Y_b M_{\tilde g}} M_h^2}{12  {\tilde M_S^4}} f_4(R)
        \right.
        \nonumber \\
        &\qquad + \left. \frac{2}{3} \frac{M_Z^2}{ {\tilde M_S^2}}
        \frac{\cos\beta \sin(\alpha + \beta)}{\sin\alpha} 
        I_3^b
        \left( f_5(R) + \frac{ {M_{\tilde g} X_b}}{ {\tilde M_S^2}} 
        f_2(R) \right) + 
        {\cal{O}} \left( \frac{m_b^2}{ {\tilde M_S^2}} \right) \right\}
        \nonumber
        \label{eq:45degexpansion}
\end{eqnarray}
 where $I_3^b=-1/2$ and the functions $f_i(R)$ are defined in ref.~\cite{HaberTemes} and 
 are normalized as $f_i(1)=1$.
 
 Notice that the first term is the dominant one in the
 limit of large $M_{SUSY}$ mass parameters and does not vanish 
 in the asymptotic limit of infinitely large 
 ${\tilde M_S}$, $M_{\tilde g}$ and $\mu$.  The second and third 
 terms are respectively  
 of ${\cal O}(M_{h^0}^2/{M^2_{SUSY}})$ and ${\cal O}(M_{Z}^2/{M^2_{SUSY}})$
  and vanish in the previous asymptotic limit. Therefore the first term gives
  the non-decoupling SUSY contribution to the $\Gamma(h^o \to b \bar b)$ partial
 width which can be of phenomenological interest. Moreover, since this term 
 is enhanced at large $\tan \beta$ it can provide important corrections to the
 branching ratio $BR(h^o \to b \bar b)$, even for a very heavy SUSY spectrum. The sign of these corrections 
 depends crucialy on the sign of $\mu M_{\tilde g}$. The previous result, when expressed in
 terms of the $h^0$ effective coupling  to $b\bar b$, agrees with the
 result in ref.~\cite{cmwpheno} based on the zero external
 momentum approximation or, equivalently, the effective Lagrangian approach.
       
 From our previous result, we conclude that there is no decoupling of sbottoms
 and gluinos  in the limit of large SUSY mass parameters for fixed 
$M_A$~\footnote{It should be 
noticed that, strictly speaking, the {\it decoupling theorem}~\cite{ac}
is not applicable to the MSSM case, since it is a theory that incorporates 
the SM chiral fermions
and the SM electroweak spontaneous symmetry breaking. For a more detailed
 discussion on this, see ref.~\cite{TesisS}}. 
  How do we then recover
 decoupling of the heavy MSSM spectra from the SM low energy physics? The
 answer to this question relies in the fact that in order to converge to SM
 predictions we need to consider not just a heavy SUSY spectra but also a heavy 
 Higgs sector. That is, besides large $M_{SUSY}$, the condition of large $M_A$ 
 is also needed. Thus, if $M_A \gg M_Z$ the light Higgs $h^0$ behaves as the SM
 Higgs boson, and the extra heavy Higgses $A$, $H^{\pm}$ and $H^0$ decouple. 
 The decoupling of SUSY particles and the extra Higgs bosons in 
 $\Delta_{SQCD}$ is seen explicitly 
 once the large $M_A$ limit of the mixing angle $\alpha$ is considered, 
  $${\cot\alpha} 
        =  {-\tan\beta} - 2 \frac{M_Z^2}{M_A^2} \tan\beta \cos 2\beta
        + {\mathcal{O}}\left(\frac{M_Z^4}{M_A^4}\right).$$
 By substituting this into our previous result we see that the non-decoupling
 terms cancel out and we get finally,
 \begin{eqnarray}
         {\Delta_{SQCD}} &
        = \frac{\alpha_s}{3\pi} 
        \left\{ \frac{2   {\mu M_{\tilde g}}} {  {\tilde M_S^2}}
        f_1(R)
            {\tan\beta} \cos 2\beta \frac{M_Z^2}{  {M_A^2}}
        - X_b\frac{  {M_{\tilde g}} M_{h^0}^2}{12 
         {\tilde M_S^4}} f_4(R)
        \right.
        \nonumber \\
        &\qquad + \left. \frac{2}{3} \frac{M_Z^2}{  {\tilde M_S^2}}
        \cos 2\beta  
        I_3^b
        \left( f_5(R) + \frac{  
        {M_{\tilde g}  {X_b}}}{  {\tilde M_S^2}} f_2(R) \right)
        + {\mathcal{O}} \left( \frac{m_b^2}{  {\tilde M_S^2}} \right) \right\}
        \nonumber
        \label{eq:45degexpansion}
\end{eqnarray}
 which clearly vanishes in the asymptotic limit of 
 $M_{\scriptscriptstyle {SUSY}}$ 
 and  $M_{\scriptscriptstyle {A}} \rightarrow \infty$.

 In conclusion, we get decoupling of the SQCD sector in 
 $h^0\rightarrow b\bar b$ decays, if and only if, both $M_{SUSY}$ and $M_A$ are 
 large.  In this limit, the dominant terms go as,
 $$\Delta_{SQCD}\sim C_1 \frac{M_Z^2}{M_A^2}+
 C_2\frac{M_{Z,h^0}^2}{M_{SUSY}^2},$$ 
 and, since both $C_1$ and $C_2$ are enhanced by $\tan\beta$, we expect
 this decoupling to be delayed for large $\tan\beta$ values. All these results are similar for the near zero mixing case,
 $\theta_{\tilde b}\sim 0^{\circ}$; for brevity we do not show these
 here (see ref.~\cite{HaberTemes}).
 
\begin{figure}[h]
\begin{center}
\epsfig{file=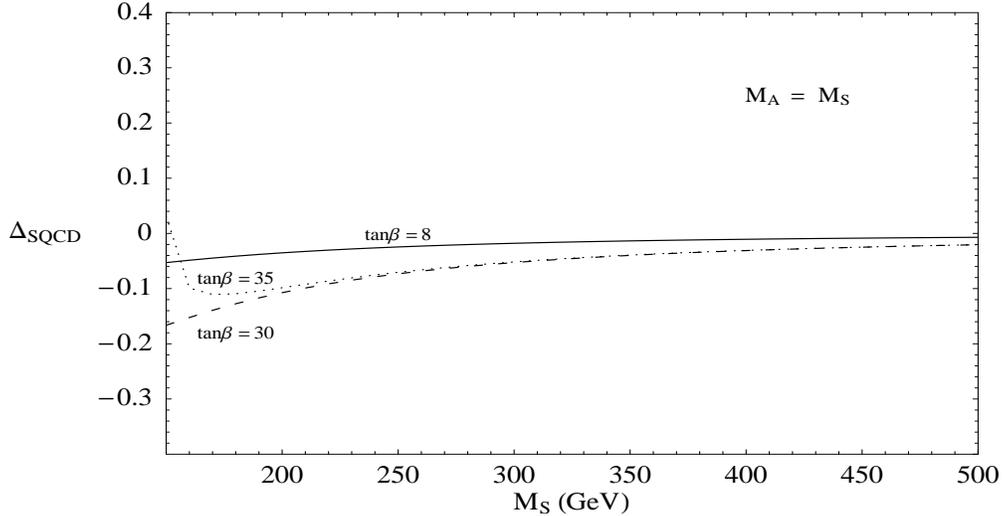,width=14cm,height=8.3cm}
\vspace{-1.2cm}
\caption[0]{Exact numerical results for $\Delta_{SQCD}$ in  
$h^0 \to b \bar b$ decay as a function of a
common MSSM scale $M_S$ and for several values of $\tan\beta$}
\label{fig:fig2}
\end{center}
\end{figure}

Finally, in order to show this decoupling numerically, we have studied a simple 
example where there is just one relevant MSSM scale, $M_S$. More specifically, 
we have chosen,
$$ M_S =M_{\tilde Q}=M_{\tilde D}=\mu=A_b=M_{\tilde g}=M_{A},$$ 
which, in the limit $M_S \gg M_Z$, gives maximal mixing, 
$\theta_{\tilde b}\sim 45^{\circ}$. In Fig.~\ref{fig:fig2} we show the 
numerical results for the exact one-loop SQCD corrections, as a function of 
this common MSSM mass scale $M_S$, and for several values of $\tan\beta$. We
can see in this figure clearly the decoupling of $\Delta_{SQCD}$ with $M_S$.
This decoupling
goes as $1/M_S^2$, in agreement with our analytical result, and is delayed for
large $\tan\beta$ values. The typical size of this correction is 
$|\Delta_{SQCD}| \leq 10 \%$ for $M_S\ge 250\,GeV$. Notice that the 
sign of $\Delta_{SQCD}$ here is
negative because of our choice of positive $\mu$ and $M_{\tilde g}$.     
 
Finally we have studied the different behavior of decoupling of squarks and
gluinos in the $\Gamma(h^0\rightarrow b\bar b)$ decay width.
 We have proved the independent decoupling of the gluinos and 
squarks whenever they are considered very heavy as
compared to the electroweak scale while keeping a large gap among their masses, that is when $M_{\tilde g} \gg M_{\tilde q} \gg M_{EW}$ or $M_{\tilde q} \gg M_{\tilde g} \gg M_{EW}$ respectively. Furthermore, the decoupling of gluinos 
is much slower than
the decoupling of squarks due to the logarithmic dependence on the 
gluino mass (see ref.~\cite{HaberTemes} for more details). 
\section{SUSY-QCD corrections to $H^+\rightarrow t\bar b$ in the decoupling 
limit}

In this section we study the SUSY-QCD corrections to the partial decay width 
$\Gamma(H^+\rightarrow t\bar b)$ at the one-loop level and to 
${\cal O}$$(\alpha_S)$. We will then analyze these corrections 
in the decoupling limit of large SUSY masses. We will present here just a short 
summary of the main numerical and analytical results, and refer the reader to 
ref.~\cite{ourHtb} for a more detailed study.

If all SUSY particles are heavy enough, $H^+$ decays dominantly into 
$t \bar b$ above the $t \bar b$ threshold. As in the case of 
$h^0\rightarrow b\bar b$, the dominant radiative corrections to 
$H^+\rightarrow t\bar b$ are the QCD corrections. At the one-loop 
level and to ${\cal O}$$(\alpha_S)$, the corresponding partial width can be 
written as,
$$\Gamma_1(H^+ \to t \bar b) \equiv \Gamma_0(H^+ \to t \bar b)
        (1 + 2   {\Delta_{QCD}} + 2   {\Delta_{SQCD}}),$$
where  $\Gamma_0(H^+ \to t \bar b)$ is the tree-level width, $\Delta_{QCD}$ 
is the correction from standard QCD, and $\Delta_{SQCD}$ is the correction from 
SUSY-QCD. The standard QCD corrections were computed in ref.~\cite{qcdhtb} and 
can be large ($+10\%$ to $-50\%$). The SUSY-QCD corrections were
        computed by using the diagrammatic approach in 
refs.~\cite{solaHTB,EberlHTB} and can be comparable or even larger than the 
standard QCD corrections in a large region of the SUSY parameter space. 

At the one-loop level and to ${\cal O}(\alpha_S) $ there are two type of diagrams that
contribute to $$\Delta_{SQCD}=\Delta_{SQCD}^{\rm loops}+
\Delta_{SQCD}^{\rm CT},$$ 
as shown in Fig.~4. 
\begin{figure}[h]
\begin{center}
\epsfig{file=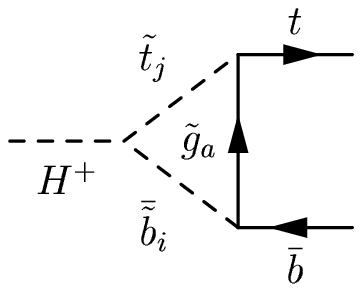}~~
\epsfig{file=AutoEn.ps}~~
\epsfig{file=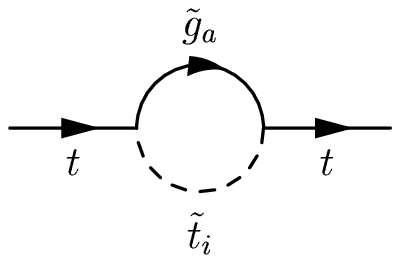}
\vspace{-1.2cm}
\caption[0]{One-loop SUSY diagrams contributing to ${\cal O}(\alpha_S)$
 to $H^+ \to t \bar b$ decay}
\end{center}
\label{fig:chargedhiggs}
\end{figure} 
The triangle diagram, with exchange of sbottoms, stops and gluinos, 
contributes to $\Delta_{SQCD}^{\rm loops}$, whereas the bottom and top
self-energy diagrams contribute to the counter-terms part 
$\Delta_{SQCD}^{\rm CT}$. The exact results in the on-shell scheme are
summarized by,

\begin{eqnarray}
&\Delta_{SQCD}^{\rm loops} =  \frac{U_t}{D}\,H_t+
\frac{U_b}{D}\,H_b\,,\nonumber \\
&\Delta_{SQCD}^{\rm CT} =  \frac{U_t}{D}\,\left(\frac{\delta m_t}{m_t}
        +\frac{1}{2}\,\delta Z_L^b+\frac{1}{2}\,\delta Z_R^t \,\right)+
 \frac{U_b}{D}\,\left( \frac{\delta m_b}{m_b}
        +\frac{1}{2}\,\delta Z_L^t+\frac{1}{2}\,\delta Z_R^b \, \right)\,,
\nonumber       
\label{eq:htbSQCD}
\end{eqnarray}
where,
\begin{eqnarray} 
D &=&
(M_{H^+}^2-m_t^2-m_b^2)\,(m_t^2\cot^2\beta+m_b^2\tan^2\beta)
-4m_t^2m_b^2\,,\nonumber\\ 
U_t &=&(M_{H^+}^2-m_t^2-m_b^2)\,m_t^2\cot^2\beta - 2m_t^2m_b^2\,,\nonumber\\ 
U_b &=&(M_{H^+}^2-m_t^2-m_b^2)\,m_b^2\tan^2\beta - 2m_t^2m_b^2\,,\nonumber\\
 H_t&=&- \frac{2\alpha _s}{3 \pi}\,\frac{G_{ab}^*}{m_t
\cot\beta} [m_t R^{(t)}_{1 b}R^{(b)*}_{1 a}(C_{11}-C_{12})+ 
m_b R^{(t)}_{2b}R^{(b)*}_{2 a}C_{12} \nonumber\\
& &+ M_{\tilde g} R^{(t)}_{2 b}R^{(b*)}_{1 a}C_0]
(m_t^2,M_{H^+}^2,M_{\tilde g}^2,M_{\tilde t_b}^2,M_{\tilde b_a}^2)\,,
\nonumber\\
H_b&=&- \frac{2\alpha _s}{3 \pi}\, \frac{G_{ab}^*}
{m_b \tan\beta} [m_t R^{(t)}_{2b}R^{(b)*}_{2 a}(C_{11}-C_{12})+ 
m_b R^{(t)}_{1 b}R^{(b)*}_{1 a}C_{12}
\nonumber\\
& &+ M_{\tilde g} R^{(t)}_{1 b}R^{(b)*}_{2 a}C_0] 
(m_t^2,M_{H^+}^2,M_{\tilde g}^2,M_{\tilde t_b}^2,M_{\tilde b_a}^2)\,,\nonumber
\label{eq:HLHR}
\end{eqnarray}
and the counter-terms are given in the on-shell scheme by,
\begin{eqnarray}
\frac{\delta m_{(t,b)}}{m_{(t,b)}}
        +\frac{1}{2}\,\delta Z_L^{(b,t)}+\frac{1}{2}\,\delta Z_R^{(t,b)} \,
        &=&
\Sigma^{(t,b)}_S(m_{(t,b)}^2)+ \frac{1}{2}\Sigma^{(t,b)}_L(m_{(t,b)}^2) -
\frac{1}{2}\Sigma^{(b,t)}_L(m_{(b,t)}^2)
\nonumber\\ 
- \frac{m_{t}^2}{2}
\left[\Sigma^{t'}_L(m_{t}^2) + \Sigma^{t'}_R(m_{t}^2)+ 2 \Sigma^{t'}_S(m_{t}^2)
\right]
&-& \frac{m_b^2}{2}
\left[\Sigma^{b'}_L(m_b^2) + \Sigma^{b'}_R(m_b^2)+ 2 \Sigma^{b'}_S(m_b^2)
\right]\nonumber 
\label{eq:CTselfE}
\end{eqnarray}
where,
\begin{eqnarray}
\Sigma^q_L(p^2)&=& - \frac{2\alpha_s }{3 \pi} 
\,|R^{(q)}_{1 a}|^2 B_1 (p^2,m^2_{\tilde{g}},m^2_{\tilde{q}_a})\,,\nonumber\\
\Sigma^q_R(p^2)&=& - \frac{2\alpha_s }{3 \pi}
\,|R^{(q)}_{2 a}|^2 B_1 (p^2,m^2_{\tilde{g}},m^2_{\tilde{q}_a})\,,\nonumber\\
\Sigma^q_S(p^2)&=& - \frac{2\alpha_s }{3 \pi}
\frac{m_{\tilde{g}}}{m_q}\, {\rm Re}
(\,R^{(q)}_{1 a}\,R^{(q)*}_{2 a}) B_0(p^2,m^2_{\tilde{g}},
m^2_{\tilde{q}_a})\,.\nonumber
\label{eq:selfs}
\end{eqnarray}
The $G_{ab}$ parametrize the $H^{+}\,\tilde b_a\,\tilde{t}_b$ couplings, and the
$R^{(q)}$ are the rotation matrices that relate the interaction-eigenstates
to the mass-eigenstates squarks. Their values in the MSSM can be found, for
instance, in ref.~\cite{ourHtb}. The above result agrees with the original
computation of refs.~\cite{solaHTB,EberlHTB}.

In order to compute $\Delta_{SQCD}$ in the decoupling limit of large SUSY 
masses, we have considered again our 'large SUSY mass limit', and we have 
performed a systematic expansion in inverse powers of the large 
SUSY mass parameters. Notice that in this case it does not make sense to 
consider the alternative limit of large $M_A$, since this parameter provides 
the charged Higgs mass value and, therefore, it must be fixed. We have obtained
analytical expansions for  $\Delta_{SQCD}$ that include up to 
${\cal O}(M_{EW}^2/ {M^2_{SUSY}})$ corrections, for all the interesting 
limiting cases of maximal and minimal mixing, in both the stop and the sbottom
sectors. For brevity, we do not present here  the complete results, which 
can be
found in ref.~\cite{ourHtb}, and we just show the most relevant result, that is, 
the dominant term in this expansion for the particular choice 
of maximal mixing. 
Thus, for  
$  \theta_{\tilde b,\tilde t} \sim 45^o$, 
$  \tilde M_S^2 \equiv 
\frac{1}{2}(M_{\tilde b_1}^2 + M_{\tilde b_2}^2)\equiv 
\frac{1}{2}(M_{\tilde t_1}^2 + M_{\tilde t_2}^2)$ and 
$R \equiv M_{\tilde g}/\tilde M_{S}$ we get:
\begin{eqnarray}
        & {\Delta_{SQCD}}
        = \frac{\alpha_s}{3\pi} 
        \left\{ \frac{-  {\mu M_{\tilde g}}}{  {\tilde M_S^2}}
        \left( \tan\beta + \cot\beta \right)
        f_1(R) + {\cal{O}} \left( \frac{M_{EW}^2}
        {  {\tilde M_S^2}} \right) \right\}
        \nonumber
        \label{eq:45degexpansion}
\end{eqnarray}
This leading term does not vanish in the heavy SUSY particle limit and,  
therefore, there is no decoupling of stops, sbottoms and gluinos in the
$\Gamma (H^+ \to t \bar b)$ decay width to one-loop level. This can be seen
clearly, for instance, for the simplest case of equal mass scales, 
$\mu=M_{\tilde g}=\tilde M_S$, where $f_1(R)=1$. This leading term, when
expressed in terms of an effective coupling of $H^+$ to $b\bar t$ is in
agreement with the previous results of refs.~\cite{CarenaDavid,EberlTodos} 
that were obtained in the zero external 
momentum 
approximation by using an effective Lagrangian approach. We see in this
result the enhancement of $\Delta_{SQCD}$ by $\tan\beta$, so that this
non-decoupling effect can be numerically important for large $\tan\beta$ 
values. As in the case of $h^0$, the sign of the SQCD correction is determined
by the sign of $M_{\tilde g}$ and $\mu$. We have obtained similar results for 
the case of minimal mixing, as can be seen in~\cite{ourHtb}.         

\begin{figure}[h]
\begin{center}
\epsfig{file=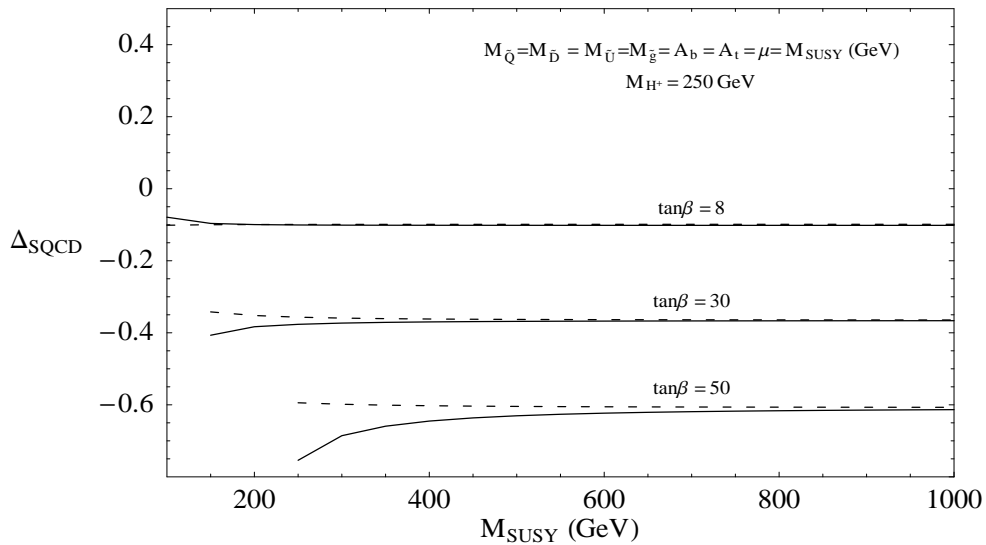,width=14cm,height=8.3cm} 
\end{center}
\vspace{-1.2cm}
\caption[0]{$\Delta_{SQCD}$ in $H^+ \rightarrow t\bar b$ decay as a function 
of the common scale $M_{SUSY}=M_S$.}
\label{fig:fig8}
\end{figure}

Finally, in order to illustrate this non-decoupling behavior 
numerically, we present in Fig.~\ref{fig:fig8} the $\Delta_{SQCD}$ 
correction as a function of a common SUSY mass scale  
$M_S = M_{\tilde Q}= M_{\tilde U}=
 M_{\tilde D}= M_{\tilde g} = A_b = A_t = \mu$. The Higgs mass has been 
 fixed to $M_{H^+}=250\,GeV$, and several values of $\tan\beta$ have been
 considered. The fact that $\Delta_{SQCD}$ tends to a non-vanishing value
 for very large $M_S$ shows precisely this non-decoupling effect. 
 The correction is quite sizeable, even for a very 
 heavy SUSY spectrum. This is particularly noticeable for large $\tan\beta$.  

As in the $h^0$ case, we have proved for the $H^+ \rightarrow t\bar b$ decay
the independent decoupling of the gluinos and 
squarks whenever they are considered very heavy as
compared to the electroweak scale and with a large gap among their masses.
 The decoupling of gluinos 
is much slower than
the decoupling of squarks due again to the logarithmic dependence on the 
gluino mass. In fact, this very slow decoupling with the gluino mass 
is the responsible for the large size of $\Delta_{SQCD}$, 
specially for large $\tan\beta$.
 For instance, if  $\tan\beta = 30$ and 
 $M_{\tilde g} = 2$ TeV we get $\Delta_{SQCD} = -40 \%$. Notice that the size 
 can be so large that the validity of the perturbative expansion can be
 questionable. We refer the reader to refs.~\cite{CarenaDavid,EberlTodos} where this subject 
 is studied and
 some techniques of resumation for a better convergence of the series are
 proposed.
\section{SUSY-QCD corrections to $ t \rightarrow W^+ b$  in the decoupling 
limit}
\begin{figure}
\begin{eqnarray*}
&&{\normalsize   \epsfig{file=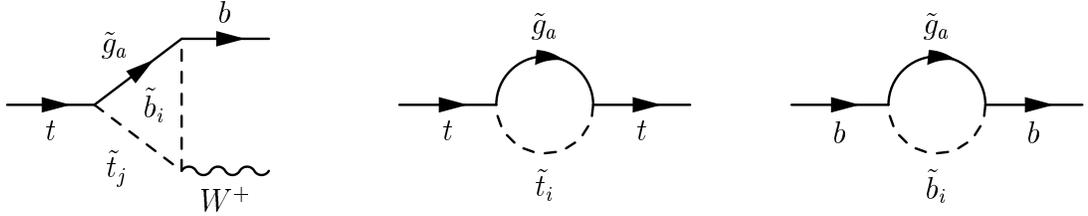}\ \ \epsfig{file=AutoEnt.ps}\ 
\ \epsfig{file=AutoEn.ps}}
\end{eqnarray*}
\vspace{-1.2cm}
\caption[0]{One-loop SUSY diagrams contributing to ${\cal O}(\alpha _S)$
in $ t \rightarrow W^+ b$ decay}
\label{fig:fig10}
\end{figure}

In this section we briefly comment on the SUSY-QCD corrections to $ t \rightarrow W^+ b$ 
at the one-loop level and to ${\cal O}(\alpha_S)$, and we study them in the
decoupling limit.  These radiative corrections 
were studied in the context of the MSSM in ref.~\cite{dhjjs}  
and are known to be important for some regions of the MSSM parameter space. 
 The standard QCD corrections are also known 
to be important and give a $\sim -10\%$ reduction in 
$\Gamma( t \rightarrow W^+ b)$~\cite{qcdtwb}. The Feynman diagrams
that contribute to the SQCD corrections are shown in 
Fig.~\ref{fig:fig10}.
The size of the SQCD corrections has been
estimated to range between $-5\%$ and $-10\%$ and are quite
insensitive to $\tan\beta$~\cite{dhjjs}. In contrast, the SUSY-Electroweak corrections that
range between $-1\%$ and $-10\%$ are known to grow with $\tan\beta$~\cite{ghjs}.
 
In order to analyze the decoupling limit in this observable we have chosen the 
simplest case with just one SUSY scale, $M_S$, which is considered very large 
as compared to the electroweak scale, $M_{EW}$, 
$$M_{\tilde Q}=M_{\tilde U}=
M_{\tilde D}=A_t=A_b=\mu=M_{\tilde g}=M_S \gg M_{EW}\,.$$

After performing an expansion of $\Delta_{SQCD}$ 
(we use here an analogous notation as in previous
sections) in inverse powers of $M_S$
we have obtained the following result for the dominant contribution,
\begin{eqnarray} 
\Delta_{SQCD} &= &-\frac{\alpha_s}{3\pi} \frac{m_t^2}{M_S^2}
\left(\frac{1}{6}+ 
\frac{1}{24}(1-\cot\beta)^2 +
\frac{1}{6} (1-\cot\beta)\right)+ {\cal{O}} \left( \frac{m_t M_W,M_W^2,
...}{M_S^2} \right)  \nonumber
\end{eqnarray} 
 From this result, we conclude that there is decoupling as $M_S$
 becomes large in the SQCD
 corrections to the 
 dominant top decay, $ t \rightarrow W^+ b$, and this decoupling which
 behaves as $(m_t^2/M_S^2)$ is not delayed for large $\tan\beta$ values. Indeed, we see in the previous 
 equation that these corrections 
 are not enhanced by $\tan\beta$ factors. Thus, we do not expect relevant 
 indirect signals from a heavy SUSY-QCD sector in this decay channel. 
\section{Indirect sensitivity to a heavy SUSY spectrum}
In this section we shortly review the main results of ref.~\cite{curiel} where
a set of optimal observables to search for indirect SUSY-QCD signals in Higgs
bosons and top decays has been proposed. These observables (for the Higgs bosons
case) are defined as ratios of Higgs branching ratios into quarks 
divided by
the corresponding Higgs branching ratios into leptons, and are  
the most sensitive to the SUSY-QCD radiative corrections. This
is a consequence from the fact that the corrections from heavy sbottoms, stops 
and gluinos {\it do not decouple} in the Higgs boson decays into quarks. We have
already shown in the previous sections some of these non-decoupling corrections
in the $h^0\rightarrow b \bar b$, and $H^+\rightarrow t \bar b$ decays.
There are non-decoupling SUSY-QCD corrections in another Higgs bosons
and top decays as well; for instance, in
$H^0\rightarrow b \bar b$ and  $A^0\rightarrow b \bar b$ decays, and in the 
top quark decay
$t \rightarrow H^+\bar b$.  In addition, all these corrections are enhanced 
by $\tan \beta$ factors and, therefore,
they can provide sizeable contributions 
to the corresponding partial decay widths 
for large enough $\tan \beta$ values, even for very 
heavy squarks and gluinos. The results of these SUSY-QCD 
non-decoupling contributions to all the relevant partial decay widths can be found in
ref.~\cite{dobado}, for arbitrary $\tan\beta$ and $M_A$ values and in 
the 'large SUSY mass limit'.

Next, we briefly comment on the proposed set of optimal observables. These are
the following~\cite{curiel}: 
\begin{eqnarray}
O_{h^o} \equiv \frac{B(h^o \rightarrow b\bar b)}{B(h^o \rightarrow \tau^+ \tau^-)}
\, \, &,& \, \, 
O_{H^o} \equiv \frac{B(H^o \rightarrow b\bar b)}{B(H^o \rightarrow \tau^+ \tau^-)}, 
\nn \\
O_{A^o} \equiv \frac{B(A^o \rightarrow b\bar b)}{B(A^o \rightarrow \tau^+ \tau^-)}
 \, \, &,& \, \, 
O_{H^+} \equiv \frac{B(H^+ \rightarrow t\bar b)}{B(H^+ \rightarrow \tau^+ \nu)}. 
\nn \\
\label{eq.interobsH}
\nn
\end{eqnarray}
and
\begin{eqnarray}
O_{t} &\equiv &\frac{B(t \to H^+ b)}{B(t \to W^+ b)}, 
\label{eq.interobst}
\nn
\end{eqnarray}
which
complements the charged Higgs observable in the low $M_{A}$ region.
These observables are specially sensitive to indirect SUSY-QCD searches 
because of the following reasons:\\
$\bullet$ The SUSY-QCD non-decoupling corrections contribute to 
the numerator but not to the denominator of these observables. 
The first will be considered 
 as the search channel. The second, as
the control channel;
 \\
$\bullet$  These corrections are
 maximized at large $\tan \beta$ and are sizeable enough as to be 
 measurable; \\
$\bullet$  The production uncertainties are minimized in ratios;  \\
$\bullet$  They will be experimentally accessible at LHC/Tevatron/Linear
Colliders; \\
$\bullet$  They will allow
 to distinguish the MSSM Higgs sector from a general 2HDM of type II. 
 
The non-decoupling SUSY-QCD contributions to these observables in the 
'large SUSY mass limit' are the following:
 
\begin{eqnarray}
O_{h^o} &= O_{h^o}^o & 
\left[ 1- \frac{2\alpha_S}{3 \pi} \frac{M_{\tilde g} \mu}{M_{\rm SUSY}^2}
(\tan \beta +\cot \alpha)  \right] 
\nn \\
O_{H^o} &= O_{H^o}^o &
\left[ 1- \frac{2\alpha_S}{3 \pi} \frac{M_{\tilde g} \mu}{M_{\rm SUSY}^2}
(\tan \beta - \tan \alpha) \right] 
\nn \\
O_{A^o} &=  O_{A^o}^o &
\left[ 1- \frac{2\alpha_S}{3 \pi} \frac{M_{\tilde g} \mu}{M_{\rm SUSY}^2}
(\tan \beta + \cot \beta) \right] 
\nn \\
O_{H^+} &= O_{H^+}^o &
\left[ 1- \frac{2\alpha_S}{3 \pi} \frac{M_{\tilde g} \mu}{M_{\rm SUSY}^2}
(\tan \beta + \cot \beta) \right]  \nn \\
O_{t}&= O_{t}^o &
\left[ 1- \frac{2\alpha_S}{3 \pi} \frac{M_{\tilde g} \mu}{M_{\rm SUSY}^2}
(\tan \beta + \cot \beta) \right]. 
\label{eq.exprobs}
\nn
\end{eqnarray}
where the leading terms, $O^o$, refer to the value of the observables 
without the SUSY particle contributions.

The numerical results for the SUSY-QCD
corrections to the set of observables $O_{h^o}$, $O_{H^o}$, $O_{A^o}$, 
$O_{H^+}$ and $O_{t}$, together with a discussion in terms of the relevant
parameters, $\tan\beta$, $M_{A}$, $\mu$ and $M_{\rm SUSY}$ 
($M_{\rm SUSY}\sim M_{\tilde q} \sim
M_{\tilde g}$), can be found in ref.~\cite{curiel}. Here we concentrate on 
the results about their sensitivity with $\tan\beta$, and compare them with 
the theoretical uncertainties that would modify the prediction of the 
observables, previous to the SUSY-QCD corrections. In particular, 
the largest theoretical uncertainties coming from the errors in $m_b$, $m_t$, 
$\alpha_s$ and from neglecting the ${\mathcal O}(\alpha^2_s)$ corrections
have been included in this analysis. 
   
\begin{figure}
\begin{center}
\epsfig{file=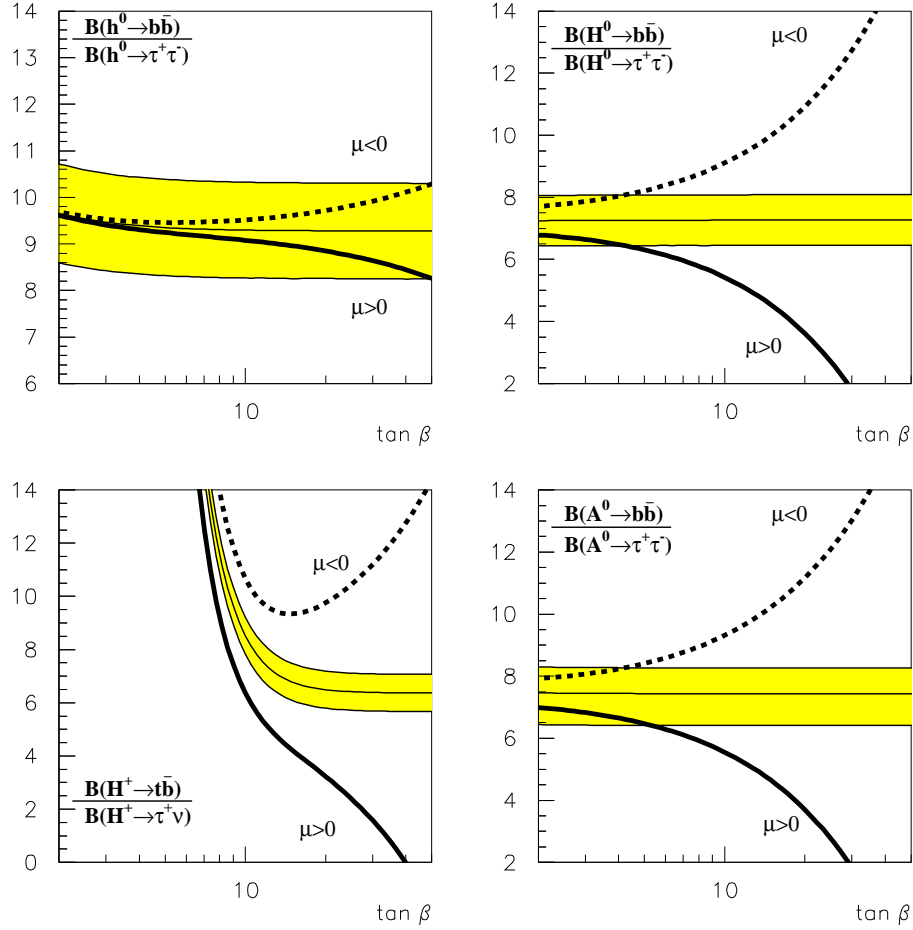,width=12cm} 
\caption{Predictions for $O_{h^o}$, $O_{H^o}$, $O_{H^+}$ and
$O_{A^o}$ as a function of
 $\tan\beta$, for $M_{A}=500$ GeV. The central lines are the corresponding
 predictions for $O^o$. The shadowed bands cover the total theoretical
 uncertainties. The bold lines represent the SUSY-QCD
corrected predictions for $M_{\rm SUSY}=M_{\tilde g}=|\mu|$.}
\label{fig.sitges500}
\end{center}
\end{figure}

\begin{figure}[h]
\begin{center}
\epsfig{file=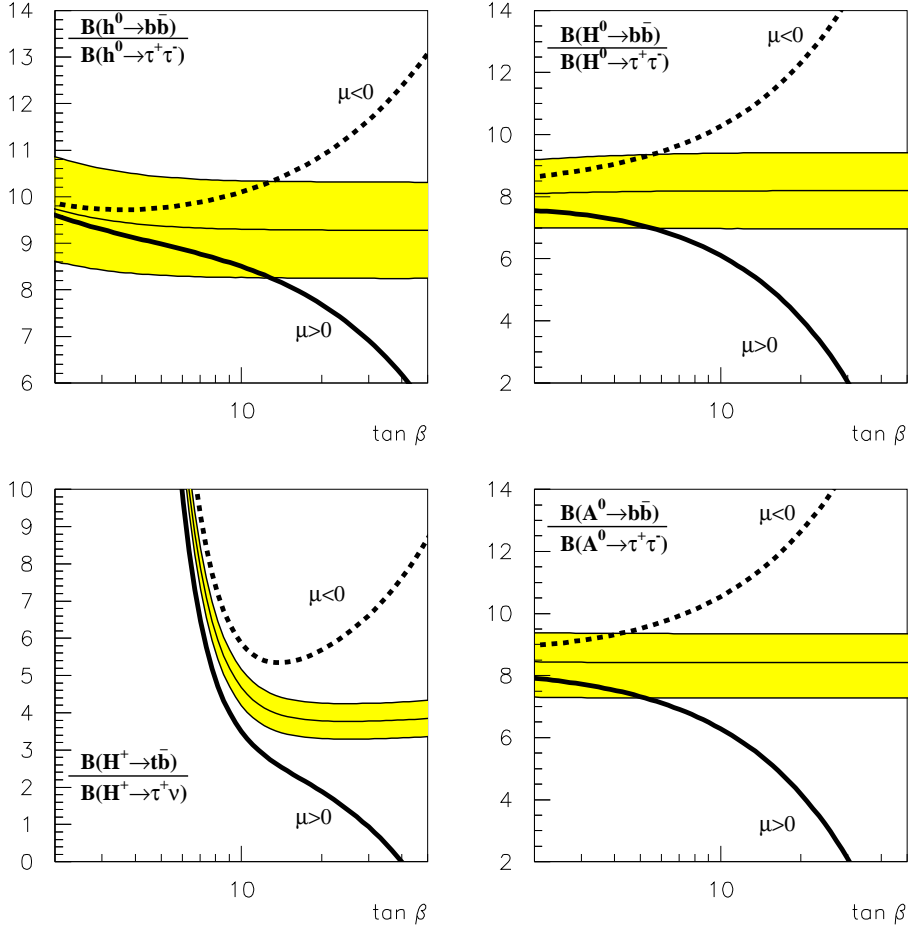,width=12cm} 
\caption{Same as in Fig.~\ref{fig.sitges500}, but for $M_{A}=250$ GeV.}
\label{fig.sitges250}
\end{center}
\end{figure}

\begin{figure}[h]
\begin{center}
\epsfig{file=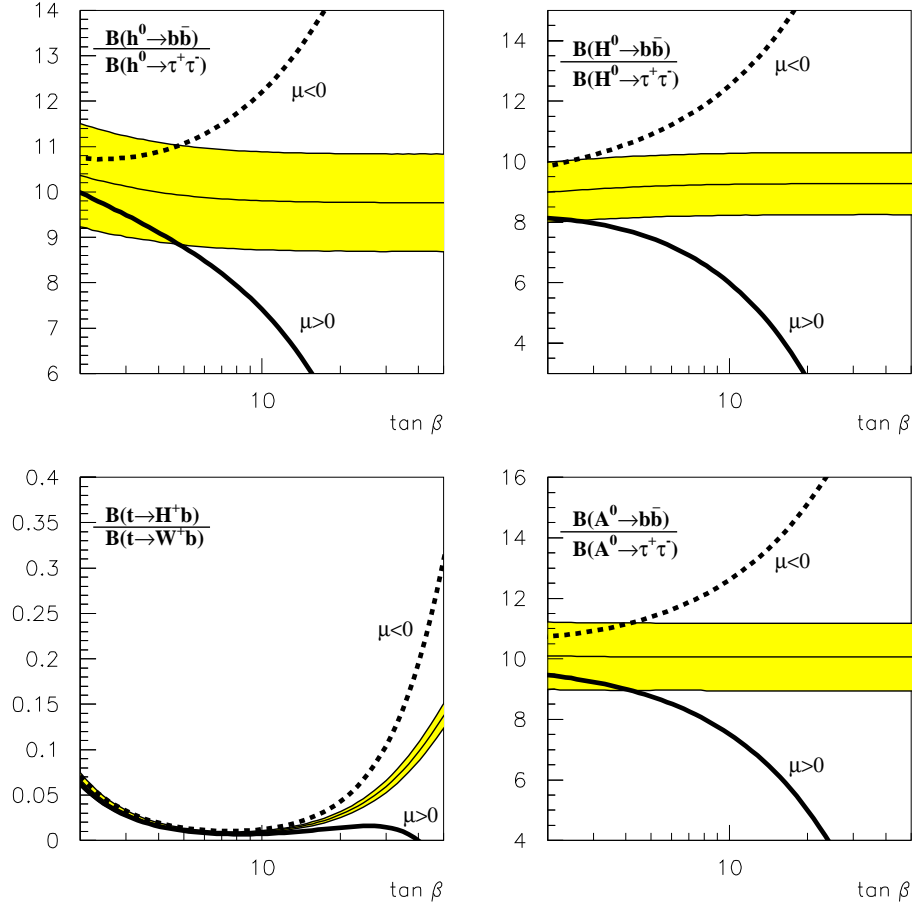,width=12cm} 
\caption{Same as in Fig.~\ref{fig.sitges500}, but for $M_{A}=100$ GeV. 
The prediction for $O_t$ appears in the lower left plot instead of 
$O_{H^+}$, since in this case $m_t > M_{H^+}+m_b$.}
\label{fig.sitges100}
\end{center}
\end{figure}

We show in ,
Figs.~\ref{fig.sitges500}-\ref{fig.sitges100} the results for 
three $M_{A}$ values in the large, medium and low $M_{A}$ region: 
$M_{A}=500$, $250$, and $100\,{\rm GeV}$ respectively. The central
lines in these figures 
follow the predictions for the observables without the SUSY-QCD contribution, 
namely, $O^o$. The corresponding total theoretical uncertainties (evaluated 
from the partial uncertainties in quadrature) are shown as shadowed
bands around the central values. The bold lines represent the SUSY-QCD
corrected predictions for $\mu > 0$ (solid), and $\mu < 0$ (dashed, respectively).
The observable $O_t$ appears in the lower left plot of Fig.~\ref{fig.sitges100} 
replacing $O_{H^+}$, since in this case $m_t > M_{H^+}+m_b$.

 One can see from the figures that, first,
the predictions for all the observables separate from the central values
as $\tan\beta$ grows. The sign of the SUSY-QCD corrections is  positive 
for $\mu<0$, and negative for $\mu>0$. The central values and their 
theoretical uncertainties 
are rather insensitive to $\tan\beta$ in the $H^o$ and $A^o$ channels. 
For the $h^o$, there is a very slight dependence at low $\tan\beta$, while 
for $H^+$ this dependence is very strong in the low $\tan\beta$ region 
and it softens for $\tan \beta>15$. 
In any case, the predictions for all the Higgs observables without the 
SUSY-QCD contributions are $\tan\beta$ insensitive in the large 
$\tan\beta$ region. This is very different for the top observable:
it depends strongly on $\tan \beta$ in all the parameter space.

Concerning the size of the SUSY-QCD corrections, we see that there
is always a $\tan\beta$ region where these are larger than the theoretical
uncertainties. For $M_{A}=500$ GeV and $M_{A}=250$ GeV, the 
observables $O_{A^o}$ and $O_{H^o}$ behave similarly and
the predictions  surpass the shadowed band 
for $\tan\beta>5$. For $M_{A}=100$ GeV, the crossing still 
happens at $\tan\beta>5$ for $A^o$, whereas for $H^o$ the 
bold lines lie outside the error band in the whole region
$2<\tan\beta<50$.
  
In the $h^o$ case the situation is qualitatively different.
For $M_{A}=500$ GeV, the SUSY-QCD correction is below the 
theoretical uncertainty for all values in the region $2<\tan\beta<50$. 
This is a manifestation of the decoupling of this correction for 
large $M_{A}$ values. 
For smaller values of $M_{A}$, as $M_{A}= 250, 100$ GeV, the 
decoupling has not effectively operated yet and the SUSY-QCD corrections 
are sizeable for large enough $\tan\beta$. In particular, for 
$M_{A}=250\ (100)$ GeV, they are larger than the theoretical error band for   
$\tan\beta>15\ (5$ respectively).
 
Regarding the charged Higgs case, two different situations must be considered. For
$M_{A}=100$ GeV, where the decay into a top and a bottom is not kinematically
allowed,  the observable $O_t$ is considered. Otherwise, for $M_{A}=250$ GeV and 
$M_{A}=500$ GeV, the relevant observable is $O_{H^+}$.  
As can be seen in Fig. \ref{fig.sitges100}, the SUSY-QCD corrections in 
$O_t$ for $M_{A}=100$ GeV are above the theoretical uncertainty for 
$\tan\beta>15$. However, the central value prediction also depends strongly on 
$\tan\beta$ 
(and $M_{A}$) and it will be difficult to identify the effect of the SUSY-QCD corrections above the
additional uncertainty related to the experimental errors on the measurement of $\tan\beta$
and $M_{A}$.
For $O_{H^+}$ and $M_{A}=500, 250$ GeV
both requirements, the correction being larger than the theoretical error and the 
leading contribution being insensitive to $\tan \beta$ are fulfilled for values larger 
than about 15.  

In summary, we have seen that these observables have a high sensitivity to
indirect SUSY-QCD signals. 

\section{Conclusions}

In this review we have analyzed the one-loop SQCD corrections to the
partial widths of $h^0 \rightarrow b \bar b$, $H^+ \rightarrow t \bar b$
and $t \rightarrow W^+ b$ decays, in the limit of large SUSY masses.  In order
to understand analytically the behavior of the SQCD corrections in this
limit,
we have performed expansions of the one-loop partial widths that are valid for 
large values of the SUSY mass parameters compared
to the electroweak scale.
We have shown that for the SUSY mass parameters and $M_A$ large and all of the
same order, the SQCD corrections in $h^0 \rightarrow b \bar b$ decay 
decouple like the inverse
square of these mass parameters, and the one-loop partial width  
$\Gamma (h^0 \rightarrow b \bar b)$ tends to its SM value. In this case 
the effective low energy theory that one obtains after integrating out
all the heavy
non-standard modes of the MSSM is precisely the SM. However, if the mass 
parameters are not
all of the same size, then this behavior is modified.  In particular, if $M_A$
is of the order of the electroweak scale, then the SQCD corrections to the 
$\Gamma (h^0 \rightarrow b \bar b)$ decay width
do not decouple in the limit of large SUSY mass parameters. We have also
presented and discussed here a similar non-decoupling SQCD correction to the  
$\Gamma (H^+ \rightarrow t \bar b)$ decay width. We have also discussed
similar SUSY non-decoupling effects that appear in 
other decay channels such as $H^0 \rightarrow b \bar b$, 
$A^0 \rightarrow b \bar b$ and $t \rightarrow H^+ b$. 

More generally, it has been shown in ref.~\cite{dobado} that, in the limit of
large SUSY mass parameters and $M_A$ of the order of the electroweak scale, 
the effective Higgs-quark-quark interactions that are generated from the
explicit integration in the path integral of 
heavy squarks and heavy gluinos, at the one-loop level,  are those of a
more general 2HDM of type III~\cite{typeIII}, where both Higgs doublets couple to both top and 
bottom-type quarks. The particular values of the corresponding effective Yukawa
couplings have also been computed~\cite{dobado}.  These non-decoupling 
SQCD 
corrections are of phenomenological interest at present 
and future colliders. In particular we have shown a recently proposed set 
of optimal observables that can provide some 
clues in the indirect search of a heavy SUSY spectrum at the next generation
colliders~\cite{curiel}.

\Acknowledgments
It is a pleasure to be contributing with this review to the special 
session at the International Meeting on Fundamental Physics dedicated to 
celebrate  Francisco J. Yndurain's sixtieth birthday and his many
contributions to Particle Physics. Paco was one of my thesis advisors and 
introduced me into many interesting topics, including some SUSY 
phenomenology. Over the years I have had numerous profitable and enjoyable 
discussions with him in very many problems and I have learnt a lot from him. 
I wish to dedicate this work on other (indirect) SUSY searches to him.
This review is entirely based on various works in collaboration with
Ana M.Curiel, Antonio Dobado, Howard E. Haber, Heather Logan, Siannah
Pe\~naranda, Stefano Rigolin, David Temes and Jorge F. de Troc\'oniz.
I wish to thank all of them for this fruitful and pleasant collaboration.        
I acknowledge the organizers of this event, Victoria Fonseca and Antonio Dobado,
for inviting me to give a talk at
this interesting and well organized conference.  
This work has been supported in part by
the Spanish Ministerio de Ciencia y Tecnologia under project 
CICYT FPA 2000-0980.   

\end{document}